\documentclass[sigconf, nonacm]{acmart}

\AtBeginDocument{%
  \providecommand\BibTeX{{%
    \normalfont B\kern-0.5em{\scshape i\kern-0.25em b}\kern-0.8em\TeX}}}

\copyrightyear{2024}
\acmYear{2024}
\setcopyright{rightsretained}
\acmConference[ICER '24 Vol. 1]{ACM Conference on International Computing
Education Research V.1}{August 13--15, 2024}{Melbourne, VIC, Australia}
\acmBooktitle{ACM Conference on International Computing Education Research
V.1 (ICER '24 Vol. 1), August 13--15, 2024, Melbourne, VIC, Australia}
\acmDOI{10.1145/3632620.3671103}
\acmISBN{979-8-4007-0475-8/24/08}
\usepackage{listings}
\usepackage{todonotes}
\usepackage{framed}
\usepackage{multirow}
\usepackage{subcaption}

\begin{document}

\title{Evaluating Contextually Personalized Programming Exercises Created with Generative AI}


\author{Evanfiya Logacheva}
\orcid{0009-0002-0962-9369}
\affiliation{
  \institution{Aalto University}
  \city{Espoo}
  \country{Finland}
}
\email{evanfiya.logacheva@aalto.fi}

\author{Arto Hellas}
\orcid{0000-0001-6502-209X}
\affiliation{
  \institution{Aalto University}
  \city{Espoo}
  \country{Finland}
}
\email{arto.hellas@aalto.fi}

\author{James Prather}
\orcid{0000-0003-2807-6042}
\affiliation{
  \institution{Abilene Christian University}
  \city{Abilene}
  \state{TX}
  \country{USA}
}
\email{james.prather@acu.edu}

\author{Sami Sarsa}
\orcid{0000-0002-7277-9282}
\affiliation{
  \institution{University of Jyväskylä}
  \city{Jyväskylä}
  \country{Finland}
}
\email{sami.j.sarsa@jyu.fi}

\author{Juho Leinonen}
\orcid{0000-0001-6829-9449}
\affiliation{
  \institution{Aalto University}
  \city{Espoo}
  \country{Finland}
}
\email{juho.2.leinonen@aalto.fi}


\begin{abstract}
Programming skills are typically developed through completing various hands-on exercises. Such programming problems can be contextualized to students' interests and cultural backgrounds. Prior research in educational psychology has demonstrated that context personalization of exercises stimulates learners' situational interests and positively affects their engagement. However, creating a varied and comprehensive set of programming exercises for students to practice on is a time-consuming and laborious task for computer science educators. Previous studies have shown that large language models can generate conceptually and contextually relevant programming exercises. Thus, they offer a possibility to automatically produce personalized programming problems to fit students' interests and needs. This article reports on a user study conducted in an elective introductory programming course that included contextually personalized programming exercises created with GPT-4. The quality of the exercises was evaluated by both the students and the authors. Additionally, this work investigated student attitudes towards the created exercises and their engagement with the system. The results demonstrate that the quality of exercises generated with GPT-4 was generally high. What is more, the course participants found them engaging and useful. This suggests that AI-generated programming problems can be a worthwhile addition to introductory programming courses, as they provide students with a practically unlimited pool of practice material tailored to their personal interests and educational needs.
\end{abstract}





\maketitle

\section{Introduction}
Learning how to program involves developing various sets of programming skills. Computing education theories present them as distinct and often sequential or hierarchical, i.e., meant to be introduced in a certain order~\cite{Malmi2023, Xie2019, Fowler2022}.
Since developing expertise in any domain requires a certain amount of training~\cite{Hambrick2020}, developing programming skills relies on practical hands-on exercises focused on particular learning objectives~\cite{Sanders2017, Programs2176, Edwards2020, Sullivan2021, Ly2021, Hausswolff2020, Mullen2017, Xie2019}. 

Research indicates that many students in computer education experience high levels of stress, frustration, confusion, and boredom, which can lead to negative outcomes~\cite{Coto2021, Hausswolff2020}. Frustration is often developed as a result of encountering various difficulties related to both coding itself and course instruction; it precedes boredom and loss of interest in subject learning~\cite{Coto2021}. Certain solutions such as substituting long weekly assignments with multiple shorter problems have been reported to alleviate students' stress and improve their performance~\cite{Programs2176}. Similarly, repetitive exercises designed for teaching students syntax in CS1 have been found effective for enhancing students' engagement and their exam scores~\cite{Edwards2020}. Such exercises have been viewed as helpful by students~\cite{Sullivan2021, Ly2021}. Overall, learning how to program hands-on has been reported to reduce students' stress levels~\cite{Hausswolff2020}. 

Programming exercises can be contextualized, i.e., worded in a narrative, or decontextualized, when they are devoid of context~\cite{Leinonen2021, Craig2017, Lovellette2017, DelCarpioGutierrez2024, Lovellette2024, Bouvier2016, Guzdial_2010}. Providing context is thought to improve students' engagement by making course materials relevant~\cite{Guzdial_2010}. Research shows mixed evidence for using contextualization in computing education. Studies using the Rainfall Problem and the Satellite Problem have found little effect on students' performance when it comes to solving contextualized vs. decontextualized programming problems~\cite{Craig2017, Lovellette2017, Lovellette2024, Bouvier2016}. However, Leinonen et al.~\cite{Leinonen2021} have discovered that context might help students avoid algebraic errors when they are tasked with programming exercises that involve mathematics. Guzdial~\cite{Guzdial2007} has reported an increase in course participant retention when contextualized exercises were introduced in course materials. In the field of educational psychology, there has been a number of studies on context personalization of mathematical problems~\cite{Bernacki2018, Walkington2018, Hoegheim2015, Hoegheim2017, Schoenherr2024}. Context personalization means tailoring learning materials to personal interests, preferences, and cultural backgrounds of learners~\cite{Bernacki2018, Walkington2017, Walkington2018, Walkington2020, Solari2022, Hoegheim2015, Hoegheim2017, Schoenherr2024}. The contextualization approach centered around students' personal interests or preferences has demonstrated positive results on situational interest and perceived utility value among students with low interest in mathematics~\cite{Hoegheim2015, Hoegheim2017}. It has also positively impacted both students whose quantitative engagement with their interests was high, in case of deep contextualization, and low, when context involved only superficial details~\cite{Walkington2018}. In a longitudinal study by Bernacki and Walkington~\cite{Bernacki2018}, context personalization has improved students' interest in mathematics and their test results. There has been no similar research on context personalization in computing education, as studies on contextualization in the domain of computer science have investigated whether it has any utility~\cite{Leinonen2021, Craig2017, Lovellette2017, Lovellette2024, Bouvier2016}. Considering possible positive impact of contextualization on students' motivation and engagement~\cite{Guzdial_2010}, Bouvier et al.~\cite{Bouvier2016} note that investing time and effort into contextualizing programming problems might be worthwhile. As they point out, developing and assessing such problems is a labor-intensive endeavor for educators, which is why there has been a shift towards researching automatic generation of programming exercises~\cite{Sarsa_2022, DelCarpioGutierrez2024}. 

Recent research on large language models (LLMs) has demonstrated promising applications of artificial intelligence in computing education~\cite{Prather_2023, Leinonen2023, Leinonen2023a, Cao_2023, Jury_2024, Liffiton2023, Wermelinger2023, MacNeil2023, Becker2023, Finnie-Ansley2022, Finnie-Ansley2023, Koutcheme2023, Koutcheme2023a}. LLMs can generate code explanations~\cite{Leinonen2023, MacNeil2023} and programming feedback~\cite{Koutcheme2023, Koutcheme2023a,hellas2023exploring,kiesler2023exploring,pankiewicz2023large}, make programming error messages more comprehensible~\cite{Leinonen2023a,santos2023always,wang2024large}, enhance intelligent tutoring systems~\cite{Cao_2023} and coding assistants~\cite{Liffiton2023}, solve programming problems~\cite{Finnie-Ansley2022, Finnie-Ansley2023, Puryear2022, Wermelinger2023}, generate worked examples~\cite{Jury_2024,hassany2024human} and programming exercises~\cite{Sarsa_2022, DelCarpioGutierrez2024,jordan2024need}. Sarsa et al.~\cite{Sarsa_2022} have shown that LLMs can be successfully used for automatically generating new exercises, containing a problem description, a sample solution, and test code. In their study, LLMs could create thematically and conceptually relevant exercises. A subsequent study by del Carpio Gutierrez et al.~\cite{DelCarpioGutierrez2024} has also demonstrated that LLMs are capable of creating high-quality exercises that include various contextual narratives in their problem descriptions and working code solutions. 

Previous research on generating novel programming problems with personalized context narratives and context personalization have inspired this work dedicated to automatic generation of programming exercises~\cite{Sarsa_2022, DelCarpioGutierrez2024, Hoegheim2015, Hoegheim2017, Bernacki2018, Bernacki2021, Walkington2017, Walkington2018, Walkington2020, Schoenherr2024}. LLMs offer a possibility to provide students with personalized exercises for hands-on practice that could alleviate existing difficulties students face when learning programming. In this work, we study the quality of programming exercises generated by LLMs, specifically GPT-4. We have both the study authors and students review the exercises. We also examine how they interact with the LLM-generated exercises in an elective introductory programming course offered by Aalto University and their feedback on the contextual personalization of the exercises. 


We seek to answer the following questions:
\begin{itemize}
    \item RQ1. How do the study authors evaluate the quality of contextually personalized exercises generated by GPT-4?
    \item RQ2. How do students evaluate the quality of contextually personalized exercises generated by GPT-4?
    \item RQ3. How do students interact with the tool that provides contextually personalized exercises?
\end{itemize}

The contributions of our work are the following:
\begin{itemize}
    \item We examine the quality of the LLM-generated programming exercises, finding that the study authors and the course participants rate them highly.
    \item We report that the course participants give overwhelmingly positive feedback on the LLM-generated programming exercises, suggesting that using them as supplementary exercises is well received by students.
    \item We study the interactions the course participants have with the generated exercises, finding that they prefer to choose the theme of the exercises over receiving a random exercise, which suggests that contextual personalization of exercises with LLMs could be an effective way to increase student engagement.
\end{itemize}

\section{Related Work}

\subsection{Context Personalization} 
Personalization in education has various applications and definitions. According to Bernacki et al.~\cite{Bernacki2021}, the most common features mentioned in its definitions are related to identifying and adapting instruction to the needs and interests of learners. However, these terms vary across educational approaches to personalization. Tetzlaff et al.~\cite{Tetzlaff2020}  contrast personalized education with traditional approaches that disregard individual characteristics of students and treat them as a unified group. 

In addition to systems that work with individualized learning needs of students, there are context personalization strategies offering learning materials according to students' out of school interests, preferences, or cultural backgrounds~\cite{Bernacki2018, Walkington2017, Walkington2018, Walkington2020, Solari2022, Hoegheim2015, Hoegheim2017, Schoenherr2024}. Solari et al.~\cite{Solari2022} summarize what constitutes learners' interests in educational psychology literature. According to them, the majority of research considers a particular type of relationship between a person and their interest, their psychological experiences and focuses on the engagement involved in a personal interest. However, they note the existing diversity of theoretical approaches to defining the object of an interest, centered around topics, domains, and practices. Solari et al.~\cite{Solari2022} propose that the personalization strategy that works with students' individual interests enhances their sense of value and meaning associated with learning. Additionally, context personalization can engage contextual grounding, which utilizes students' prior experiences with a particular context~\cite{Bernacki2018}. It is thought to ease long-term memory recollection, lessen the probability of conceptual errors, and facilitate understanding of subject domain concepts~\cite{Bernacki2018}.  

It has also been suggested that context personalization may stimulate students' situational interest, which is characterized by improved engagement and attention caused by conspicuous parts of learners' environments~\cite{Bernacki2018, Walkington2018, Hoegheim2015, Hoegheim2017, hidi2006, Michaelis2022}. Situational interest, in turn, can lead to the development of individual interest in a subject domain, for example, mathematics or computing~\cite{Hoegheim2015, Bernacki2018, hidi2006, Michaelis2022}. Hidi and Renninger~\cite{hidi2006} suggest a four-phase model of situational and individual interest development. The first phase is triggered situational interest, which may serve as a precursor to the inclination to engage with particular content over a period of time; the second phase, maintained situational interest, is characterized by focused attention and may nudge learners towards further engagement in a more advanced phase. Individual interest, comprised of emerging and well-developed (well-established) phases, is typically self-generated, while situational interest is often externally supported~\cite{hidi2006}. Nevertheless, emerging individual interest might be hindered by lack of support, resources, and positive reinforcement, whereas well-established interest is stable throughout time~\cite{Michaelis2022}. In computing education, developing individual interest is thought to rely on consistent positive situational interest built over time, along with discovering value and knowledge in educational content and building ``a sense of belonging'' to it~\cite{Michaelis2022}. Thus, contextually personalized learning materials could trigger the initial stage of situational interest development and help learners develop a sense of attachment and value in educational content.  

Context personalization approaches differ in terms of the depth of implementation. Surface personalization involves simple references to students' interests, while deep personalization draws on students' prior knowledge and uses it in learning materials~\cite{Walkington2017, Walkington2018, Walkington2020}. Personalization also diverges at the level of granularity~\cite{Walkington2020}. At a fine grain size, personalized problems focus on particular topics, for example, a student's favorite ice hockey teams. In contrast, when they center around domains, e.g., sports, personalization is applied at a broader level of granularity. Additionally, context personalization strategies can be based on either interests or preferences~\cite{Hoegheim2017}. Here, preferences are options from students' mundane life aspects, for instance, friendships and food, while interests pertain to objects of sustained engagement, e.g., music and sports~\cite{Hoegheim2017}. Context personalization is also used as an adaption of problems to real-life contexts experienced by students without involving their personal interests. Schoenherr~\cite{Schoenherr2024} used locations and objects of students' hometown for personalizing problems.

While a series of empirical studies on context personalization has been conducted on children in mathematics classes or after-school programs~\cite{Bernacki2018, Walkington2018, Hoegheim2015, Hoegheim2017, Schoenherr2024}, there has been little attention dedicated to it in computing education.  Høgheim and Reber~\cite{Hoegheim2015} found that in mathematics, surface context personalization positively affected situational interest and perceived utility value among students whose subject area interest was low. Their subsequent study showed that personalization based on individual preferences mostly benefited learners with low interest and perceived competence in mathematics~\cite{Hoegheim2017}. The authors pointed out that the level of granularity of personalization could have an effect on situational interest. Similarly, Walkington and Bernacki~\cite{Walkington2018} discovered that the depth of personalization influenced its effectiveness. Learners who were less involved with quantitative functions in their interests were more likely to be positively affected by surface personalization, while those whose quantitative engagement with their interests was higher benefited more from deep personalization. Bernacki and Walkington~\cite{Bernacki2018} reported an improvement in developing students' interest in mathematics through context personalization in a longitudinal study. They also noted a beneficial influence of solving personalized problems on students' test performance. In a study conducted as an after-school mathematics program, using familiar locations in context personalization was found to increase students' intrinsic (interest and enjoyment) and attainment (related to personal or identity-based importance) values~\cite{Schoenherr2024, ECCLES2020101859}.

Contextually personalized problems have a potential to solve a number of issues associated with frustration and boredom experienced by students in programming education~\cite{Coto2021}. They are, nevertheless, time-consuming to create~\cite{Bouvier2016}. However, as LLMs display state-of-the-art results in many natural language processing problems and even code generation ~\cite{Li_2022, Koutcheme2023, Koutcheme2023a, Tian2023, Finnie-Ansley2022, Finnie-Ansley2023, Puryear2022, Wermelinger2023, Chen2021EvaluatingLL}, creating contextually personalized programming exercises might be achieved with their help.

\subsection{Perceptions of Assessment Quality}



The way students and teachers perceive assessment is relevant to this study as we generate uncredited, additional practice opportunities for students for formative assessment utilizing generative AI. Struyven et al.~\cite{struyven2003students} conducted a literature review of 36 empirical studies. They found that, in general, students find assessment positive (i.e., beneficial for their learning) and fair (i.e., accurately and justly measuring their progress towards their learning goals) if the assessment relates to authentic tasks, presents reasonable demands, encourages them to apply knowledge to realistic contexts, focuses on the need to develop a range of skills, and is perceived to have long-term benefits~\cite{struyven2003students, sambell1997but}.

This is supported by later empirical findings. Van Dinther et al.~\cite{van2014student} studied the perceptions of assessment of 138 first-year elementary teacher students at a Dutch university. They focused on studying the links between perceptions, self-efficacy, and performance. They found that formative assessment where students create ``a quality product or observable performance in a real-life situation'' and where feedback is tied to the task and criteria increases self-efficacy, which in turn is likely to lead to more learning. They also argue that the presence of sufficient practice is a requirement for mastery of the topic.

Gerritsen et al.~\cite{gerritsen2019students} studied the perceptions of 204 higher education students based on six aspects of assessment quality: effects on learning, fairness, conditions, interpretation of test scores, authenticity, and credibility. They found that students who had more positive perceptions of the effects of assessment on learning were more likely to employ deeper and strategic learning approaches, whereas students who had negative perceptions were more likely to apply a surface learning approach, which has been linked to worse learning outcomes~\cite{marton1976qualitative}. They argue that this is due to students deepening their approach if they find the assessment appropriately challenging and motivating. Similarly, Gulikers et al.~\cite{gulikers2006relations} found that the more authentic students find the tasks that they are solving, the deeper the study approach they choose, which should result in enhanced learning.

Related to teachers' perceptions of assessment quality, Sach~\cite{sach2012teachers} analyzed the perceptions of 67 lower and middle school teachers of assessment. She found that teachers value formative assessment and believe it to enhance learning. However, teachers were less confident in actually employing formative assessment practices in their own courses.

These prior works suggest that it is imperative to make (formative) assessment tasks authentic~\cite{struyven2003students,van2014student,gulikers2006relations}, contextually relevant~\cite{struyven2003students}, and motivating~\cite{gerritsen2019students}. Especially for the last two aspects, utilizing LLMs to personalize the context of the tasks to the interests of the student could be useful. Similarly, it is important to provide students with enough opportunities for practice in order for them to achieve mastery of the topic~\cite{van2014student}, which could be scaffolded by utilizing LLMs to generate programming exercises for formative assessment.
\subsection{Large Language Models in Computing Education}
Large language models (LLMs) have exhibited outstanding performance in many tasks, including code generation~\cite{Li_2022, Koutcheme2023, Koutcheme2023a, Tian2023, Finnie-Ansley2022, Finnie-Ansley2023, Puryear2022, Wermelinger2023, Chen2021EvaluatingLL}. Recent advances have prompted active research on the use of LLMs in computing education~\cite{Prather_2023}. There has been a number of studies focused on LLMs' ability to solve programming problems~\cite{Finnie-Ansley2022, Finnie-Ansley2023, Puryear2022, Wermelinger2023}. Finnie-Ansley et al. discovered that even early versions of these models such as the now-deprecated Codex perform better than students in both CS1 and CS2 programming courses when it comes to code writing~\cite{Finnie-Ansley2022, Finnie-Ansley2023}. Puryear and Sprint~\cite{Puryear2022} and Wermelinger~\cite{Wermelinger2023} have demonstrated that Github Copilot can generate coding solutions similar to those written by students. In addition to solving programming exercises and tests, LLMs have been found to successfully generate code explanations~\cite{Sarsa_2022, Leinonen2023, MacNeil2023}. AI-generated code explanations were seen as helpful by learners~\cite{MacNeil2023} and clearer and more accurate than those created by students themselves~\cite{Leinonen2023}.

Codex has also been assessed for its ability to generate explanations for programming error messages~\cite{Leinonen2023a}. Although the results of the study by Leinonen et al.~\cite{Leinonen2023a} were mixed, the authors have noted that such explanations could be used as a scaffold for understanding programming errors. Later works by Santos et al.~\cite{santos2023always} and Wang et al.~\cite{wang2024large} that used more recent models have demonstrated better performance. Santos et al.~\cite{santos2023always} found that providing the LLM the code that produced the error helped to enhance error messages. Wang et al.~\cite{wang2024large} discovered that students who utilized GPT-4-enhanced error messages repeated errors less frequently and required fewer attempts to fix them compared to those students who received traditional error messages.

LLMs have also been used to develop coding assistants for students, for instance, to power an on-demand tool providing support in undergraduate courses~\cite{Liffiton2023}. Liffiton et al.~\cite{Liffiton2023} argue that such tools are valuable as they offer instant help to students when they cannot get in touch with course instructors. According to the authors, an online system like theirs can also relieve students' anxiety about reaching out for help. Generative AI has also been investigated for generation of other sorts of learning materials, for example, worked examples. Jury et al.~\cite{Jury_2024} have successfully implemented a tool for creating interactive worked examples. As the authors point out, developing such examples is time-consuming and LLMs can significantly facilitate their creation. Furthermore, GPT-3 has been used to boost gamification in an intelligent tutoring system~\cite{Cao_2023}. The system was offered to a group of Chinese students in the UK with an intent to make their learning environment more inclusive and increase their sense of belonging. It was well received by the students who reported feeling supported~\cite{Cao_2023}.

\subsection{Automatic Programming Exercise Generation}
Prior to the emergence of modern generative AI models based on LLMs, template-based approaches were popular for automatic generation of programming exercises~\cite{prados2005automatic,wakatani_automatic_2015,wakatani_evaluation_2016,zavala_use_2018}. In these template-based, parameterized exercises, certain parts of their problem descriptions are parameterized, meaning that specific parts of an exercise are modified according to defined parameters. This technically allows one to generate an infinite number of variations where, for example, numbers or specific statements in problem descriptions are different. As each student is presented with minor exercise differences, one typical use case for such exercises is to prevent plagiarism ~\cite{radosevic_automatic_2010}.

Using context free grammar to form program templates~\cite{ade2019syntactic,peess2023grammar} and generating random exercises from various models such as UML diagrams or mathematical notation~\cite{sovietov2021automatic,tiam2018procedural} have been suggested to improve purely parameter-based template approaches. These methods allow a way to individualize exercises for students. Such template-based approaches also make it possible to effectively personalize exercise content to themes and topics, as demonstrated by Zavala and Mendoza~\cite{zavala_use_2018}, who contextualized the generated exercises though linked open data. Nonetheless, they fail to generate completely new and varied exercises with deep personalization without extensive manual effort.


As LLMs displayed state-of-the-art performance in code generation and explanation tasks, Sarsa et al.~\cite{Sarsa_2022} investigated the use of OpenAI Codex for generating programming exercise task descriptions with model solutions and automated tests. Their study yielded promising results, as the generated exercises were mostly sensible and sometimes of sufficient quality to be handed to students to solve without modification. However, roughly one fifth of the generated exercises did not make sense, and Codex sometimes struggled to accommodate concepts given as keywords to personalize the exercises. The authors additionally noted that Codex managed poorly in creating sample solutions and automated tests. On the other hand, focusing solely on solution and test generation, Chen et al.~\cite{chen2022codet} achieved much better results by generating multiple solution and test pairs and then picking a sample where the generated solution passed the generated tests.

Replicating the work of Sarsa et al.~\cite{Sarsa_2022} with a more modern AI model, namely GPT-4, del Carpio Gutierrez et al.~\cite{DelCarpioGutierrez2024} evaluated generated exercises for context relevance, description clarity, and problem sensibility in a larger study involving multiple prompting strategies. They assessed the generated content using rubrics and found the quality of the content to be high. Another replication of the work by Sarsa et al.~\cite{Sarsa_2022} authored by Jordan et al.~\cite{jordan2024need} explored the performance of GPT-3.5 in generating exercises in four natural languages. They found that problems generated in English, Spanish, and Vietnamese were mostly accurate and understandable and would only require minor modifications before giving them to students. However, the quality of the exercises generated in Tamil was poor, indicating that current models still do not completely generalize across natural languages.

In a similar work with a more manual approach, Speth et al.~\cite{speth2023investigating} generated programming exercise sheets with ChatGPT (GPT-3.5) chat sessions by having an instructor provide ideas and context to the model and then iteratively leveraging ChatGPT to refine the generated exercises within the ChatGPT session. While they noticed that ChatGPT was adept at quickly generating good exercises, they noted that instructors nearly always resorted to minor manual edits to improve exercise quality. Phung et al.~\cite{phung2023generative} explored ChatGPT and GPT-4 against human tutors in generating debugging quizzes that could help students practice specific concepts (among other things). They focused on creating simplified versions of students' buggy pieces of code to help them practice solving bugs that they encounter, whereas our aim is to create new practice exercises contextualized to various themes and topics. 

The unique contribution of our work is the focus on evaluating context personalization specifically. In addition, only one prior work on automatic exercise generation using LLMs actually had students complete the generated exercises -- the study by Speth et al.~\cite{speth2023investigating} -- and thus, more evidence on how students perceive LLM-generated exercises is needed.


\section{Methods}
\subsection{Prompt Engineering}
The prompt used for generating the final exercise set went through multiple revisions both in terms of wording the prompt and the role of the request and determining an appropriate temperature value, which regulates randomness of the model's responses. The original version of the prompt only asked the model to produce a new exercise based on the example we provided and to contextualize it to a particular topic. This, however, resulted in the model often producing the same response when asked for an exercise contextualized to a popular theme, e.g., pets or outdoor activities. For example, when asking for an exercise related to pets, the model would almost solely generate exercises related to cats and dogs. To circumvent this limitation, we asked the model to list popular entities or things within certain categories, which constituted the themes used for personalization. The top level themes were arbitrarily chosen by the authors, whereas the topics within the themes were generated by the model. For instance, to obtain a list of topics for the \textit{pets} theme, we asked the model to generate a list of ten popular pets. When asked to generate popular handicrafts, GPT-4 included home-brewing, which is why we had to explicitly forbid it from mentioning alcohol in order to keep the exercises suitable for a wide audience. Additionally, we tried to make the topics contextually relevant to our users by using such keywords as ``Finland'' and ``the Nordic countries'' in our requests. Our themes were the following: \textit{outdoor activities,	literature,	historical landmarks,	classical music,	pop music,	cartoons,	food,	pets,	sports,	video games,	nature destinations,	handicrafts,	art,	Christmas,	party games,	board games}. Additionally, we had to ask the model to omit words related to diseases, as it included \textit{Pandemic} in the \textit{board game} theme, which we considered unacceptable following the COVID-19 pandemic.

Once the model was given the topics to create new exercises, other issues appeared. When asked to generate a new problem about guinea pigs, it produced a response that contained incorrect arithmetical calculations and mentions of weight gain. Following this occurrence, we made a decision to restrict the model further and ask it to exclude trigger words associated with mental or physical disorders in order to keep the course materials inoffensive. The response related to guinea pigs was as follows.

\begin{quote}
\textit{    'title': 'Guinea Pig Weight Gain', 'problemDescription': 'Write a program that calculates the weight gain of a guinea pig. The program should ask the user for the initial weight of the guinea pig and the amount of food it has eaten in grams. Then, it should calculate and print the weight gain of the guinea pig. If the user enters the initial weight as 500 grams and the amount of food eaten as 200 grams, the program should print the weight gain as 200 grams. Similarly, if the user enters the initial weight as 400 grams and the amount of food eaten as 300 grams, the program should print the weight gain as -100 grams.}
\end{quote}

When determining an appropriate value for the temperature parameter, we gave the model the same prompt asking it to generate a new exercise about Mickey Mouse with two different temperature values. When the temperature value was set to 0.7, it produced the somewhat humorous problem description found below.

\begin{quote}
\textit{'title': 'Mickey Mouse Age Calculator', 'problemDescription': 'Mickey Mouse wants to know his age in dog years. Write a program that asks the user for Mickey Mouse's age in human years, and then calculates and prints his age in dog years. The conversion rate is 1 human year equals 7 dog years. The program should work as follows:```10 70```'}
\end{quote}

Lowering the temperature to 0.3 resulted in it completely ignoring the topic and producing an exercise unrelated to Mickey Mouse.
\begin{quote}
\textit{'title': 'Mickey Mouse Age Calculator', 'problemDescription': 'Write a program that asks the user for their birth year and the current year, and then calculates and prints their age. The program should work as follows:```Enter your birth year: 1990 Enter the current year: 2022 Your age is: 32```'}
\end{quote}

\begin{figure*}[ht]
\centering
\includegraphics[width=0.8\textwidth]{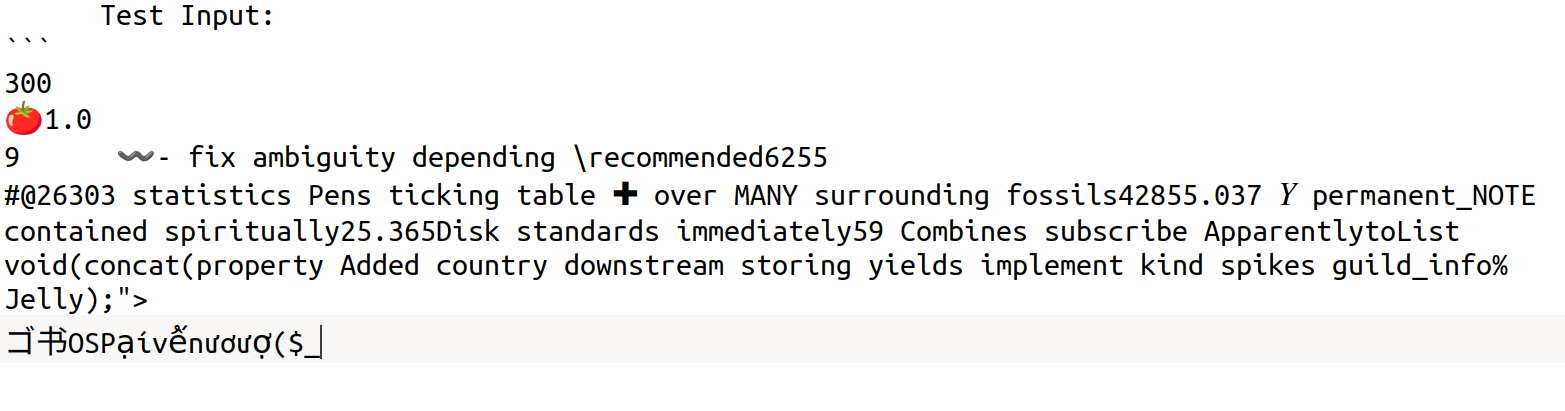}
\caption{A screenshot of nonsensical output containing emojis.}
\label{fig:emoji_output}
\end{figure*}

Raising the temperature to 1.5 led to the model hallucinating. Figure \ref{fig:emoji_output} contains a response with emojis.

Similarly, the temperature value of 2.0 resulted in another absurd response. Eventually, the temperature value of 0.5 was chosen to keep the model restrained to the structure of the examples we provided but allow for some variation in its output. 

Additionally, the model was asked to include certain concepts covered in the first three chapters of the course. Two concepts were introduced in each chapter, and the model was instructed to use the concepts covered in the current and previous chapters. Without this restraint, the model would often produce exercises unrelated to the concepts discussed in the course. The first chapter covered \textit{user input} and \textit{program output}, the second one introduced \textit{variables} and \textit{arithmetics}, and the third one was dedicated to \textit{conditional
statements} and \textit{logical operators}. To generate \textit{normal} difficulty exercises, we asked the model to create new problems at the same difficulty level as our examples. To make them a bit more difficult, we requested slightly more complex exercises. The model was instructed to avoid loops due to the limitations of the tool, which could potentially become stuck in an endless loop.

The final prompt is as follows (the variables in bold). 

\begin{quote}
\textit{Please generate a short programming exercise in Dart based on the example that I will provide. It should be about \textbf{\$theme}, specifically \textbf{\$topic}. It should be at the same difficulty level as the example \textbf{/or} It should be slightly more complex than the example. It should mainly cover \textbf{\$concept1} but can also include \textbf{\$concept2}. Please follow the structure of the example and stay within its scope. You are allowed to include the following concepts in the new exercise: \textbf{\$concepts}. Do not use loops. Your response should be a JSON string. Here is the example: \textbf{\$example\_exercise}}
\end{quote}

In order to provide blanket specifications to the model, we experimented with the role that is included in the GPT-4 request.
The role mentioned that our students were programming novices, which was necessary so that the model would not produce exercises too advanced for an introductory course. We also repeated the structure of our desired output to prevent the model from deviating from it. To prevent copyright issues, the model was asked to avoid citing lyrics or literary works since our topics included music and book authors. Finally, as our students were not native English speakers, we asked the model to stick to simple English.
Our final role included numerous instructions that the model was to follow in its output:
\begin{quote}
\textit{I want you to act as a programming teacher for an introductory Dart course. Your students are programming novices. I will provide some coding example exercises, and it will be your job to invent new ones. They should contain the following name-value pairs in JSON: title, problemDescription, exampleSolution, starterCode, tests. Your responses should be written in simple English. Do not cite music lyrics or books. Do not include any greetings, be concise. Do not mention trigger words associated with mental or physical disorders, for example, weight loss or diet.}
\end{quote}

\subsection{Exercise Generation}
For the purposes of this study, we chose to pregenerate the exercises instead of having an on-demand system. This was done for multiple reasons. First, pregenerating the exercises allowed us to check them for material that could be offensive to a wide audience. This is a concern as LLM-generated content might include biases that were present in the training data of the LLM~\cite{Ferrara_2023}. Second, possible outages in the availability of the LLM API could have caused technical disruptions in the system when students wanted to use the tool. Lastly, during prompt engineering we noticed that the LLM would sometimes generate faulty code (see Table~\ref{tab:gen_exer}), and in many cases this would have made exercises impossible to assess automatically (as they require working unit tests generated by the LLM). 

The exercise generation was done with the aim to obtain a varied set of problems for each chapter covering different concepts. To achieve the desired variety, we originally created three exercises for each combination of difficulty (\textit{normal} and \textit{advanced}), concept (two concepts per chapter, six in total)\footnote{user input, program output, variables, arithmetics, conditional
statements, logical operators}, and theme (each theme contained 10 topics to choose from). We randomly picked three topics to generate novel exercises, and some were kept as spare in case the model failed to generate an exercise with working code for a particular combination. Originally, each chapter had 96 exercises\footnote{8 themes per chapter $\times$ 3 exercises per theme $\times$ 2 difficulty levels $\times$ 2 concepts $=$ 96 exercises per chapter}; however, some of them contained faulty test code -- 25.0\% from the first chapter, 20.8\% from the second, 92.7\% from the third (the detailed summary can be found in Table \ref{tab:gen_exer}). The astounding number of broken exercises in the third chapter was caused by  an extra space in the example exercise that was fed to the model and replicated in the generated instances. The issue was fixed manually. Some exercises had an issue with missing escape characters, which led to the final pool consisting of 93 exercises in the first chapter and 95 exercises in the remaining two chapters each. Although we experimented with explicitly asking the model to remember to use the escape character, it did not. After multiple requests, it seemed unlikely it would generate working code, which is why we stopped the exercise generation. 
\begin{table}
\centering
\caption{Summary of exercise generation per chapter.}
\label{tab:gen_exer}
\begin{tabular}{|c|p{13em}|p{7em}|} \hline 

 \textbf{Chapter}& \textbf{Percentage of exercises with faulty code} & \textbf{Final number of exercises} \\ \hline 

   1 & 25.0\% &  93 \\ \hline 
   2 & 20.8\%& 95 \\ \hline 
   3 & 92.7\%& 95 \\ \hline

\end{tabular}
\end{table}

Although there were 16 themes used for exercise generation in total, some of them were closely related, for example, \textit{sports} and \textit{outdoor activities}. Due to this, we split the themes into three subsets. The first chapter contained \textit{Christmas, classical music, food, historical landmarks, literature, party games, video games},  and \textit{outdoor activities}. The second had \textit{art, board games, cartoons, handicrafts, nature destinations, pets, pop music, sports} for the offered themes. The third chapter contained the mixture of the themes from the previous chapters: \textit{literature, pop music, video games, party games, outdoor activities, handicrafts, arts, pets.} For each theme there were ten specific topics generated with the help of the model that were used in the prompt.

\subsection{Study Context}

The study was conducted in an open online introductory programming course that uses the Dart programming language and is offered by Aalto University in Finland. The course is worth 2 ECTS credit points, which corresponds to approximately 50 study hours. The course covers input and output, variables and arithmetics, conditionals and logical operators, looping, functions, and lists and maps. The course uses an online textbook platform. The online textbook has intertwined embedded theory and exercise parts with programming exercises and quizzes, where the exercises are automatically assessed by the platform. The programming exercises are completed in an embedded online integrated development environment (IDE), which is opened through the platform when students work on programming exercises. The IDE comes with normal IDE functionality, including syntax and error highlighting and the possibility to run the programs within the IDE; programming exercises are also submitted through the IDE.  

As the course is an open elective online course, the course participants include both Aalto University students and lifelong learners. The platform does not distinguish between them, and as is usual for open online courses, attrition rates have room for improvement. Out of the students who complete at least one exercise in the first chapter of the course, 51\% continue to the second chapter, and 44\% continue to the third chapter.

The responsible teacher of the course did not provide access to demographic data for this study. However, when considering the demographic data, prior research on the course has highlighted that the participants come from a range of backgrounds~\cite{sarsa2022who}. Most of the participants who continue past the first chapters of the course are between 26-35 or 36-55 years old, have some experience from tertiary education, have taken no prior programming courses, participate in the course due to being interested in the topic, and self-estimate their programming knowledge as very low. The vast majority of the course population are lifelong learners as CS majors at our university have other mandatory introductory programming courses. Of those identifying their gender, approximately half of the participants self-identify as men, a bit more than one third as women, and the remaining either as other than men or women or do not wish to disclose their gender. 



\subsection{Tool}

For the purposes of the study, we implemented a new component to the platform that allows retrieving LLM-generated programming exercises and showing them in the embedded IDE. When retrieving an exercise, students can select a theme, a concept, and a difficulty level. They can also allow that any of these are chosen randomly. Based on the selection, the platform then retrieves a problem description and starter code, which are shown to the student. Once the student has completed the exercise, they can submit the exercise for grading in a similar way as if they worked  on the standard course exercises. The LLM-generated exercises were evaluated using automated tests that were also generated using an LLM.

The component was embedded into the first three chapters of the course: (1) input and output, (2) variables and arithmetics, and (3) conditionals and logical operators. The component was shown at the end of each chapter. The instructions before the component that the students saw were as follows (translated from Finnish).

\begin{quote}
\textit{The first chapters of the course offer AI-generated programming exercises. You can try these exercises below. Please choose a theme, a concept, and difficulty. After this, click the ``Get Exercise''-button.}\vspace{1mm}

\textit{At this point, the course platform will load a problem description and the exercise and show a programming environment where you can work on the exercise. Once you finish the exercise, we will ask you for feedback about the programming exercise.}\vspace{1mm}

\textit{You can complete as many AI-generated programming exercises as you want. The exercises and problem descriptions are currently only available in English.}
\end{quote}

Completing the exercises did not yield any course points, and the students were not compensated in any other form. The tool had a progress bar that highlighted progress in the AI-generated exercises in each chapter. For each chapter, the progress bar filled after completing three exercises, at which point the students also saw a greyed out trophy icon being colored as yellow. This slight gamification was added to try nudge students into completing at least some of the AI-generated exercises.

Figure~\ref{fig:exercise-selection} below shows a screenshot of the component with the selection of the theme, the concept, and the difficulty, as well as a retrieved problem description.

\begin{figure*}[ht]
\centering
\includegraphics[width=0.8\textwidth]{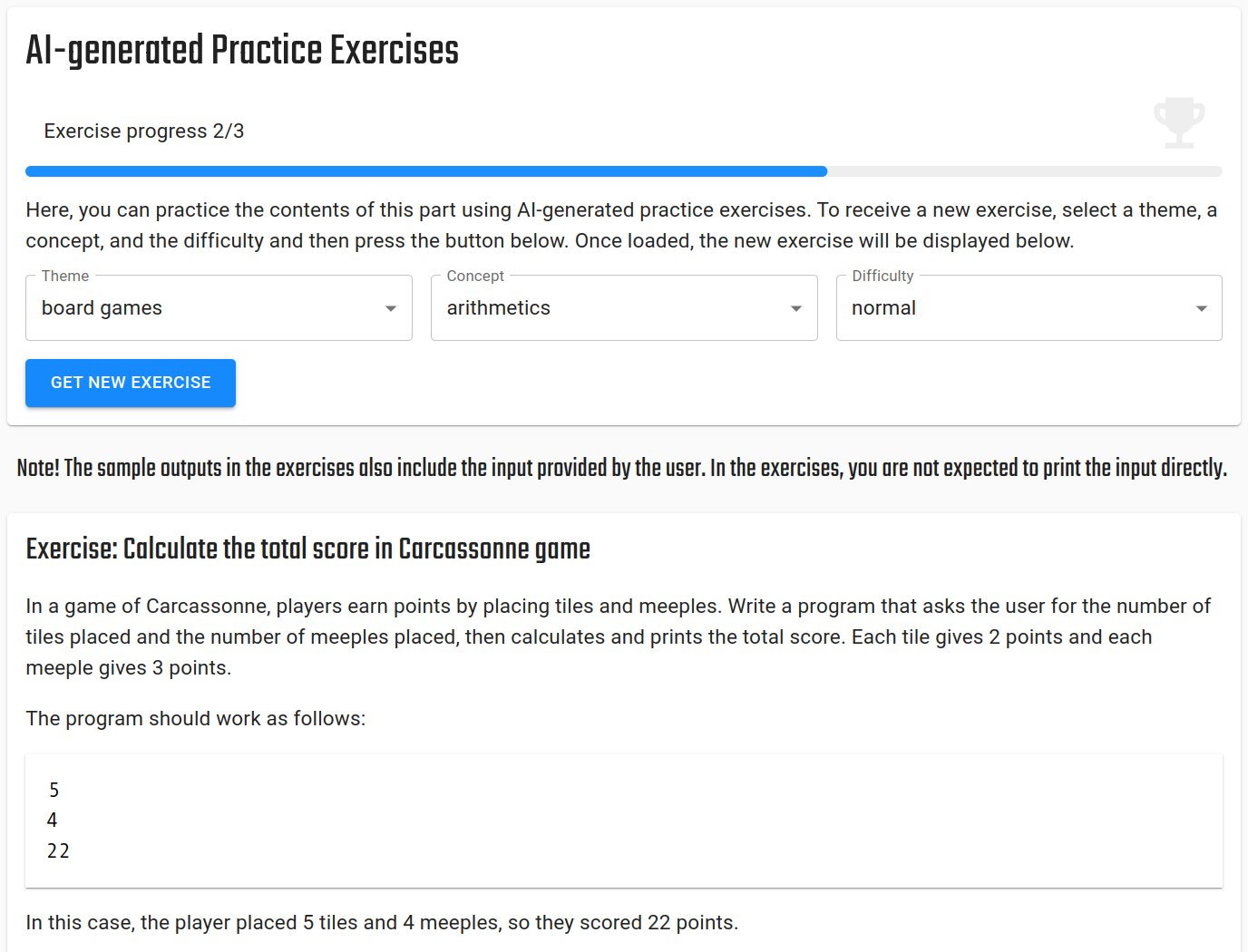}
\caption{A screenshot of the exercise selection functionality with a problem description shown. In the screenshot, the user has already completed two exercises. The user has selected ``board games'' as the theme, ``arithmetics'' as the concept, and ``normal'' as the difficulty, and then pressed the ``Get Exercise''-button. The button is labeled as ``Get New Exercise'' as an exercise has already been retrieved. In this example, an exercise about Carcassonne has been retrieved.}
\label{fig:exercise-selection}
\end{figure*}

\subsection{Data Collection}
\label{sec:data_collection}

The platform collected data on fetching exercises, where the data included a student identifier, a timestamp, the selections (theme, concept, difficulty), and the retrieved exercise. In addition, the platform collected data on the submissions, where the data similarly included the student identifier, a timestamp, an identifier for the exercise, and the submitted code. We also implemented survey functionality to the component. Once a student completed an exercise, they were shown a survey regarding the exercise, with the following four questions.

\begin{enumerate}
\item The exercise description was clear.
\item The exercise description matched the selected theme.
\item The exercise description matched the selected concept.
\item The exercise difficulty matched the selected difficulty.
\end{enumerate}

Once a student completed three exercises within a chapter, they were shown a different survey with the following questions. 

\begin{enumerate}
\item The exercises were useful for my learning.
\item The exercises were engaging to me.
\item The theme selection was satisfactory to me.
\item I enjoyed being able to select themes that match my interests.
\item Please provide open feedback on the AI-generated practice exercises.
\end{enumerate}

With the exception of the last question (``Please provide open feedback on the AI-generated practice exercises.''), the questions were given as 7-item Likert-scale questions with the following prompt: \textit{Please indicate how much you agree with each of the following statements, or how true it is about you. Use the scale from 1 to 7, where 1 is `Strongly disagree' and 7 is `Strongly agree'.}

Responding to the surveys was optional -- students could skip the surveys if they wished. Data for this study was collected over four weeks in March 2024.

\subsection{Approach}

\subsubsection{RQ1: Expert Evaluation by the Study Authors}
For research question 1, \textit{``How do the study authors evaluate the quality of contextually personalized exercises generated by GPT-4?''}, we follow the methods of prior work that has evaluated LLM-generated exercises by developing a rubric for quality assessment of exercises~\cite{Sarsa_2022,DelCarpioGutierrez2024,jordan2024need}. The rubric questions can be found in Table~\ref{tab:exp_eval_irr}.

Out of the 283 programming exercises, 33 were randomly selected for inter-rater reliability analysis. The five authors discussed the rubric and rated the 33 exercises according to the rubric. The results of the inter-rater reliability analysis are presented in Table~\ref{tab:exp_eval_irr}. We report both the percentage agreement and Gwet's AC1 statistic. We chose Gwet's AC1 as the inter-rater reliability metric due to the limitations of other multi-rater IRR statistics, such as Krippendorf's alpha and Fleiss' kappa, caused by high agreement between raters ~\cite{gwet2008computing,gwet2014handbook,falotico2015fleiss,feinstein1990high}.

After the IRR analysis, each author individually evaluated a set of 50 randomly selected exercises. Thus, all the exercises that were included in the tool were assessed in the expert evaluation. We report the exact numbers of answers for each question as well as percentages.

For the final set of evaluations, for the 33 exercises that were evaluated in the IRR analysis, we use the majority vote value. There were only four ties for majority value, all of them occurring for the ``shallow vs. deep personalization'' question. For three ties, they were between ``unsure'' and ``deep'' and for one, between ``unsure'' and ``shallow''. In these four cases, we chose the other option than ``unsure'' for the final evaluation set.


\subsubsection{RQ2: Student Evaluation}
For research question 2, \textit{``How do students evaluate the quality of contextually personalized exercises generated by GPT-4?''}, we analyze the feedback on the generated exercises provided by the course participants.

We report the distribution of the Likert-answers to the survey questions, which are listed in Section~\ref{sec:data_collection}. Due to the low number of responses to the open feedback question (n = 4), we do not analyze it.



\subsubsection{RQ3: Student Interactions}
For research question 3, \textit{``How do students interact with the tool that provides contextually personalized exercises?''}, we report how many exercises were fetched and solved in addition to the number of the course participants who solved all three exercises necessary for obtaining a trophy. We analyzed what choices they made when fetching exercises in terms of the theme and difficulty selection. Since their activity levels differed, we calculated theme popularity as well as their' preference for random theme selection as an average ratio for the normalized number of exercises retrieved per student (i.e., for each user, we calculated how often they selected a specific vs. a random theme as a percentage). We additionally report how much time passed between each exercise retrieval and its solution being submitted per exercise. This is done to describe the impact of exercise difficulty levels on student performance as measured by time-on-task, which has been shown to strongly correlate with performance~\cite{leinonen2022time}. Using the expert evaluation of the exercises, we analyze how unclear exercise descriptions, inclusion of advanced concepts, and inadequate difficulty correlated with student performance.

\section{Results}







\subsection{RQ1: Expert Evaluation by the Study Authors}


\begin{table*}[h!]
\centering
\caption{Percentage agreement and Gwet's AC1 for expert evaluation.}
\label{tab:exp_eval_irr}
\begin{tabular}{|l|l|c|c|} \hline 

\textbf{Question} & \textbf{Answer Options} & \textbf{\% Agreement} & \textbf{Gwet's AC1} \\ \hline  

1. The exercise description was clear & Yes/Partially/No & 79\% & 0.90 \\ \hline  
2. The exercise description matched the selected theme & Yes/Partially/No & 94\% & 0.97 \\ \hline  
3. The exercise description matched the selected topic & Yes/Partially/No & 85\% & 0.94 \\ \hline  
4. The exercise description matched the selected concept & Yes/No & 100\% & 1.00 \\ \hline  
5. Included concepts that were too advanced & Yes/No & 94\% & 0.96 \\ \hline  
6. The exercise difficulty matched the selected difficulty & Too easy/Okay/Too difficult & 27\% & 0.49 \\ \hline  
7. Shallow vs. deep personalization & Deep/Unsure/Shallow & 18\% & 0.34 \\ \hline 

\end{tabular}
\end{table*}

The results of the inter-rater reliability analysis for the expert evaluation are presented in Table~\ref{tab:exp_eval_irr}. From the table, it can be seen that the level of agreement varied between the questions. Agreement was quite high for the first five questions but was lower for the last two.

\begin{table*}[h!]
\centering
\caption{Summary of expert evaluation.}
\label{tab:expert_eval_results}
\begin{tabular}{|l|l|c|c|}
\hline
\textbf{Question} & \textbf{Response} & \textbf{Count} & \textbf{Percentage} \\ \hline
1. The exercise description was clear & Yes & 273 & 96.5\% \\ \cline{2-4}
 & Partially & 10 & 3.5\% \\ \cline{2-4}
 & No & 0 & 0.0\% \\ \hline
2. The exercise description matched the selected theme & Yes & 272 & 96.1\% \\ \cline{2-4}
 & Partially & 7 & 2.5\% \\ \cline{2-4}
 & No & 4 & 1.4\% \\ \hline
3. The exercise description matched the selected topic & Yes & 270 & 95.4\% \\ \cline{2-4}
 & Partially & 9 & 3.2\% \\ \cline{2-4}
 & No & 4 & 1.4\% \\ \hline
4. The exercise description matched the selected concept & Yes & 248 & 87.6\% \\ \cline{2-4}
 & No & 35 & 12.4\% \\ \hline
5. Included concepts that were too advanced & Yes & 14 & 4.9\% \\ \cline{2-4}
 & No & 269 & 95.1\% \\ \hline
6. The exercise difficulty matched the selected difficulty & Too easy & 112 & 39.6\% \\ \cline{2-4}
 & Okay & 154 & 54.4\% \\ \cline{2-4}
 & Too difficult & 17 & 6.0\% \\ \hline
7. Shallow vs. deep personalization & Deep & 75 & 26.5\% \\ \cline{2-4}
 & Unsure & 27 & 9.5\% \\ \cline{2-4}
 & Shallow & 181 & 64.0\% \\ \hline
\end{tabular}
\end{table*}

The results of the expert evaluation of the 283 generated exercises is shown in Table~\ref{tab:expert_eval_results}. Answering the first question, \textit{``The exercise description was clear''}, we considered the overwhelming majority of the exercises to be  clear. As for the second, \textit{``The exercise description matched the selected theme''}, and third, \textit{``The exercise description matched the selected topic''}, questions, we found that almost all of them matched both their theme and topic. When evaluating the exercises for the fourth question,\textit{``The exercise description matched the selected concept''}, we concluded that 87.6\% of them corresponded to their requested concept. When it comes to assessing whether the exercises included concepts that were too advanced (the fifth question), we found that most of them did not contain concepts that were out of scope for the corresponding chapter of the course. However, when evaluating whether the exercise difficulty matched the selected difficulty (the sixth question), we concluded that the difficulty of the exercises was frequently not satisfactory. We found that the difficulty was rated as ``too easy'' in 39.6\% of the cases, ``too difficult'' in 6.0\%, and ``okay'' in 54.4\% of the exercises. As for the depth of personalization (the seventh question), it was shallow in the majority of the generated exercises (64\%), while the rest were either somewhere in between (9.5\%) or deeply personalized (26.5\%). 

The cases where there were too difficult concepts included occurred mostly in the first chapter. In these cases, when the model was asked for an ``advanced'' exercise related to user input or output, it would produce an exercise requiring conditionals (introduced in Chapter 3), e.g., print different outputs depending on user input. See Figure~\ref{example:tol} for an example where the concept was ``user input'' (Chapter 1), but the exercise required conditionals to solve.

\begin{figure}[h!]
\centering
\caption{An exercise example with overly advanced concepts.}
\label{example:tol}
\begin{framed}
\raggedright
Write a program that asks the user for their favorite historical landmark. If the user's favorite is the Tower of London, the program should print a message to the user ‘Tower of London is a great choice!’, where Tower of London is the landmark entered by the user. For example, with the input `Tower of London', the program output is as follows:

\vspace{\baselineskip}
\begin{small}
\begin{verbatim}
```
What is your favorite historical landmark?
Tower of London
Tower of London is a great choice!
```
\end{verbatim}
\end{small}
However, if the user enters a different landmark, the program should simply print the name of that landmark.
\end{framed}
\end{figure}

Related to matching the theme and topic, we found that sometimes the problem description would match them but contain some factual errors. See Figure~\ref{example:pictionary} for an example. In the example, it is claimed that in Pictionary, teams score between 0 and 3 points depending on whether their guess is correct, almost correct, wrong, or not guessing at all. In actuality, teams just score a single point for a correct guess\footnote{As per the rules: \url{https://service.mattel.com/instruction_sheets/T5132-0920.pdf}}.

\begin{figure}[h!]
\centering
\caption{An exercise containing a factual error.}
\label{example:pictionary}
\begin{framed}
\raggedright
In a game of Pictionary, each team gets a score between 0 and 3 for each round, based on the following scale: 
\begin{small}
\begin{verbatim}
<table>
<tr>
<th>Score</th>
<th>Result</th>
</tr>
<tr>
<th>3</th>
<th>Correct Guess</th>
</tr>
<tr>
<th>2</th>
<th>Almost Correct</th>
</tr>
<tr>
<th>1</th>
<th>Wrong Guess</th>
</tr>
<tr>
<th>0</th>
<th>No Guess</th>
</tr>
</table>
\end{verbatim}
\end{small}
Write a program that asks the user for a score and prints the result related to that score. If the user enters any other score, the program should print the message <code>Invalid Score!</code>.

\vspace{\baselineskip}
Below is an example of the expected operation of the program.
\vspace{\baselineskip}
\begin{small}
\begin{verbatim}
<pre>
What score?
<b>&lt; 2</b>
Almost Correct
</pre>

Another example.

<pre>
What score?
<b>&lt; 4</b>
Invalid Score!
</pre>
\end{verbatim}
\end{small}
\end{framed}
\end{figure}

As mentioned above, most of the exercises were only shallowly personalized. Often, there was a sentence or two about the theme or topic in the problem description, but the actual exercise was not directly relevant to the context. See Figure~\ref{example:ariana} for an example of a shallowly personalized exercise. The theme for the exercise was ``pop music'', and its topic was ``Ariana Grande''. However, only the first sentence of the exercise mentions Ariana Grande and album sales, while the rest of it is not related to the theme or topic. There were some deeply personalized exercises too (for example, Figure~\ref{example:picasso}). In the exercise, most of the problem description deals with the theme (``art'') and the topic (``Pablo Picasso''), and the task is directly relevant to the topic.

\begin{figure}[h!]
\centering
\caption{An exercise with shallow personalization.}
\label{example:ariana}
\begin{framed}
\raggedright
Write a program that asks the user for the number of albums Ariana Grande sold in two different years. Then, print the difference between them. If the user enters the numbers 3 million and 2 million, the program should print the number 1 million. Similarly, if the user enters the numbers 2 million and 3 million, the program should print the number -1 million.

\vspace{\baselineskip}
The program should work as follows:
\vspace{\baselineskip}
\begin{small}
\begin{verbatim}
```
3
2
1
```
```
2
3
-1
```
\end{verbatim}
\end{small}
\end{framed}
\end{figure}

\begin{figure}[h!]
\centering
\caption{An exercise with deep personalization.}
\label{example:picasso}
\begin{framed}
\raggedright
Pablo Picasso had different periods in his art career. One of them is the Blue Period from 1901 to 1904. During this period, he painted essentially monochromatic paintings in shades of blue and blue-green. Write a program that asks the user for a year and prints whether or not it falls within Picasso's Blue Period. If the user enters any other year, the program should print the message <code>Not in the Blue Period!</code>.

\vspace{\baselineskip}
Below is an example of the expected operation of the program.
\vspace{\baselineskip}
\begin{small}
\begin{verbatim}
<pre>
Which year?
<b>&lt; 1902</b>
Yes, in the Blue Period.
</pre>
\end{verbatim}
\end{small}
\vspace{\baselineskip}
Another example.
\vspace{\baselineskip}
\begin{small}
\begin{verbatim}
<pre>
Which year?
<b>&lt; 1900</b>
Not in the Blue Period!
</pre>
\end{verbatim}
\end{small}
\end{framed}
\end{figure}

There were a few cases where the problem description matched the topic (which was not visible to the course participants) but not the theme (which was chosen by the student). See Figure~\ref{example:cookies} for an example. Here, the theme was ``Christmas'', and the topic was ``baking cookies''. The exercise is clearly relevant to the topic but not directly relevant to the theme (and would probably better fit the theme of ``Cooking'', for example).

\begin{figure}[h!]
\centering
\caption{An exercise relevant to the topic but not to the theme.}
\label{example:cookies}
\begin{framed}
\raggedright
Write a program that asks the user for the number of cookies they want to bake. After this, the program prints a message to the user `You will bake N cookies!', where N is the number entered by the user. For example, with the input `5', the program output is as follows:
\vspace{\baselineskip}
\begin{small}
\begin{verbatim}
```
How many cookies do you want to bake?
5
You will bake 5 cookies!
```    
\end{verbatim}
\end{small}
\vspace{\baselineskip}
Similarly, if the user enters the number `12', the program output is as follows:
\vspace{\baselineskip}
\begin{small}
\begin{verbatim}
```
How many cookies do you want to bake?
12
You will bake 12 cookies!
```
\end{verbatim}
\end{small}
\end{framed}
\end{figure}

In many cases, particularly for the arithmetic exercises, their problem descriptions contained deeply personalized details, but the corresponding code solutions were strikingly similar to the example problems we provided to the model. For instance, the solution to the problem in Figure \ref{example:handball} could be solved with exactly the same code containing a subtraction operation as the example exercise (Figure \ref{example:example_exercise}) given to the model. However, the exercise could be considered deeply personalized as calculating the difference in goals is relevant to the theme (``sports'') and the topic (``handball''). 

\begin{figure}[h!]
\centering
\caption{An exercise that can be solved using the same code as the exercise that was provided to the model as an example.}
\label{example:handball}
\begin{framed}
\raggedright
Write a program that asks the user for the number of goals scored by two different teams in a handball match, and then prints the difference between them. If team A scored 10 goals and team B scored 3 goals, the program should print the number 7. Similarly, if team A scored 4 goals and team B scored 8 goals, the program should print the number -4.

\vspace{\baselineskip}
The program should work as follows:
\vspace{\baselineskip}
\begin{small}
\begin{verbatim}
```
12
15
-3
```
\end{verbatim}
\end{small}
\end{framed}
\end{figure}

\begin{figure}[h!]
\centering
\caption{One of the three example exercises used as input.}
\label{example:example_exercise}
\begin{framed}
\raggedright
Write a program that asks the user for two numbers, and then prints the difference between them. If the user enters the numbers 10 and 3, the program should print the number 7. Similarly, if the user enters the numbers 4 and 8, the program should print the number -4.

\vspace{\baselineskip}
The program should work as follows:
\vspace{\baselineskip}
\begin{small}
\begin{verbatim}
```
12
15
-3
```
\end{verbatim}
\end{small}
\end{framed}
\end{figure}

\subsection{RQ2: Student Evaluation}



\begin{figure*}
\centering
\begin{subfigure}[b]{0.9\textwidth}
   \includegraphics[width=1\linewidth]{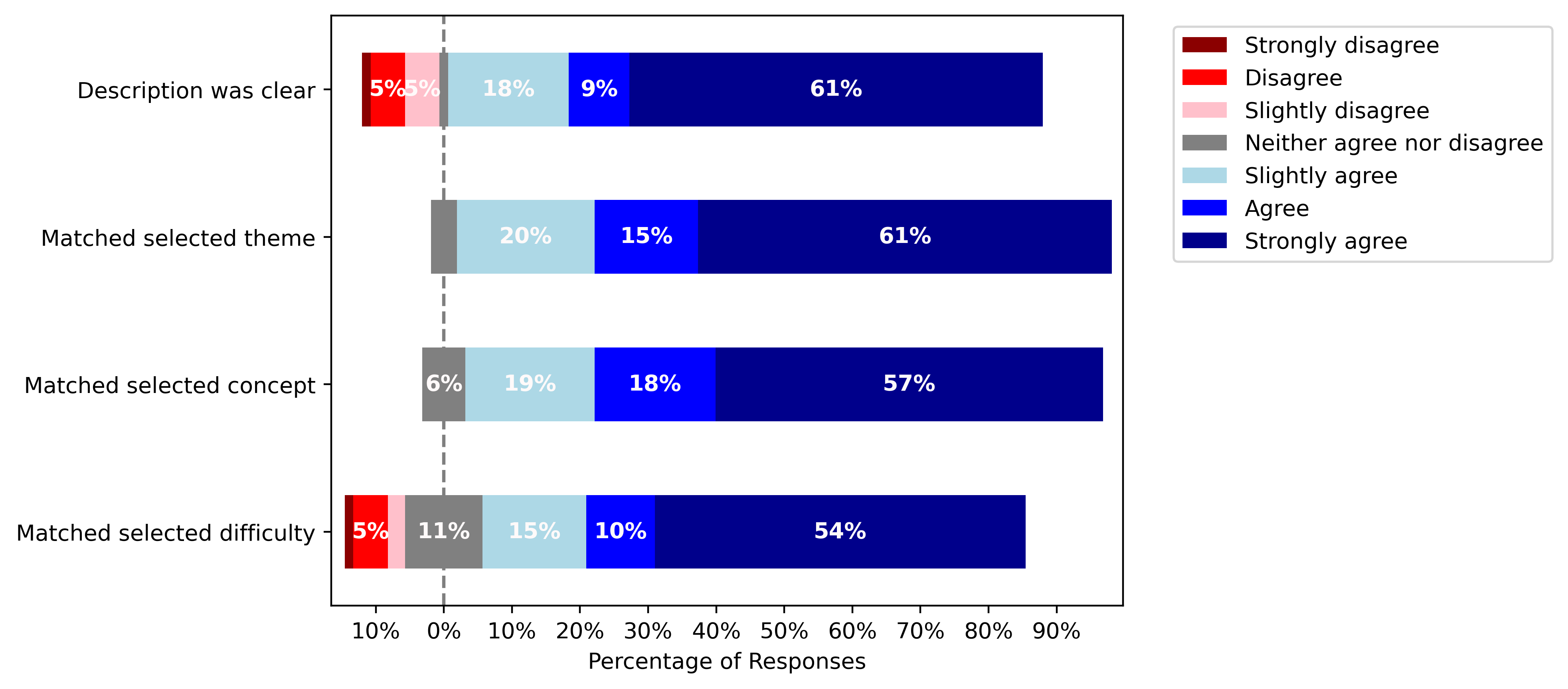}
   \caption{Student evaluation results (per exercise).}
   \label{fig:after_each_exercise} 
\end{subfigure}

\begin{subfigure}[b]{0.9\textwidth}
   \includegraphics[width=1\linewidth]{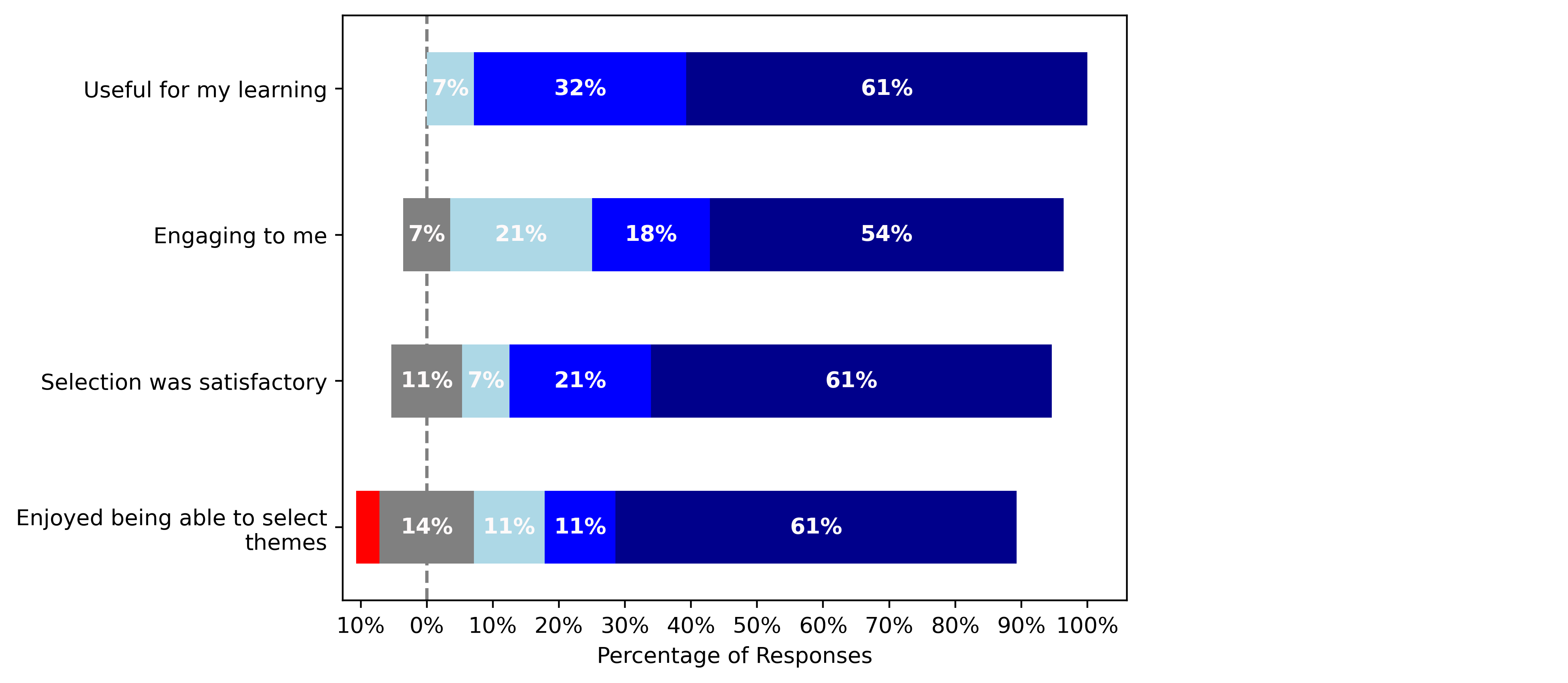}
   \caption{Student evaluation results (after three exercises).}
   \label{fig:after_three_exercises}
\end{subfigure}

\caption{Student evaluation results. (a) shows the responses to the questionnaire that was shown after each exercise and (b) shows the responses to the questionnaire that was shown after three completed exercises.}
\label{fig:student_evaluation_results} 
\end{figure*}

Figure~\ref{fig:student_evaluation_results} shows the results of the student evaluation of the generated exercises. Figure~\ref{fig:after_each_exercise} shows the distribution of the Likert-scale responses to the survey given after each exercise and Figure~\ref{fig:after_three_exercises} shows the distribution for the survey shown to the students after they had completed three exercises. There were a total of 79 responses to the exercise-specific survey and 28 responses to the general survey shown after completing three exercises.

From the figures, it can be seen that the student feedback was overwhelmingly positive, with over half of the students strongly agreeing with every statement. The five statements that did not receive any disagreements are the following: \textit{matched selected theme, matched selected concept, useful for my learning, engaging to me,} and \textit{theme selection was satisfactory}. The three statements that received negative responses (\textit{description was clear, matched selected difficulty,} and \textit{enjoyed being able to select themes}) only had a few responses disagreeing with the statements (12\%, 10\%, and 3\% respectively).

\subsection{RQ3: Student Interactions}

Tables \ref{tab:summary2} and~\ref{tab:summary_tool} contain summaries of descriptive statistics for the interaction data collected by the tool. Table~\ref{tab:summary2} shows the overall number of the students who fetched and solved exercises and the range, mean, median, and standard deviation for the fetched and solved exercises. Table~\ref{tab:summary_tool} shows the number of the fetched and solved exercises and the number of the individual course participants split by chapter. Note that the same person might have used the tool across multiple chapters, which is why the number summed in Table \ref{tab:summary_tool} is higher than the number of the students in Table \ref{tab:summary2}. A total of thirty seven users solved three or more exercises, another two solved two exercises, and the remaining eight submitted one correct solution.

\begin{table}
\centering
\caption{Summary of student interaction data (for those students who fetched at least one exercise).}
\label{tab:summary2}
\begin{tabular}{|l|l|l|l|l|l|} \hline 

 & \textbf{N} & \textbf{Range} & \textbf{Mean} & \textbf{Median} & \textbf{SD} \\ \hline 

   Fetched & 68 &  [1, 27] &     5.32 &     3 &     5.55 \\ \hline 
   Solved & 47 & [0, 10] &     2.87 &     3 &     2.88 \\ \hline

\end{tabular}
\end{table}

\begin{table*}[h]
\centering
\caption{Summary of student interactions with the tool split by chapter.}
\label{tab:summary_tool}
\begin{tabular}{|c|c|c|c|c|} \hline 

\textbf{Chapter} & \textbf{\# fetched exercises} & \textbf{\# solved exercises}  & \textbf{\# users who retrieved exercises} & \textbf{\# users who solved exercises} \\ \hline 

1& 177 & 106 & 53 & 40 \\ \hline 
2& 108 & 60 & 29 & 22\\ \hline 
3& 71 & 28& 20 & 11\\ \hline

\end{tabular}
\end{table*}

\begin{figure*}[ht]
\centering
\includegraphics[width=0.8\textwidth]{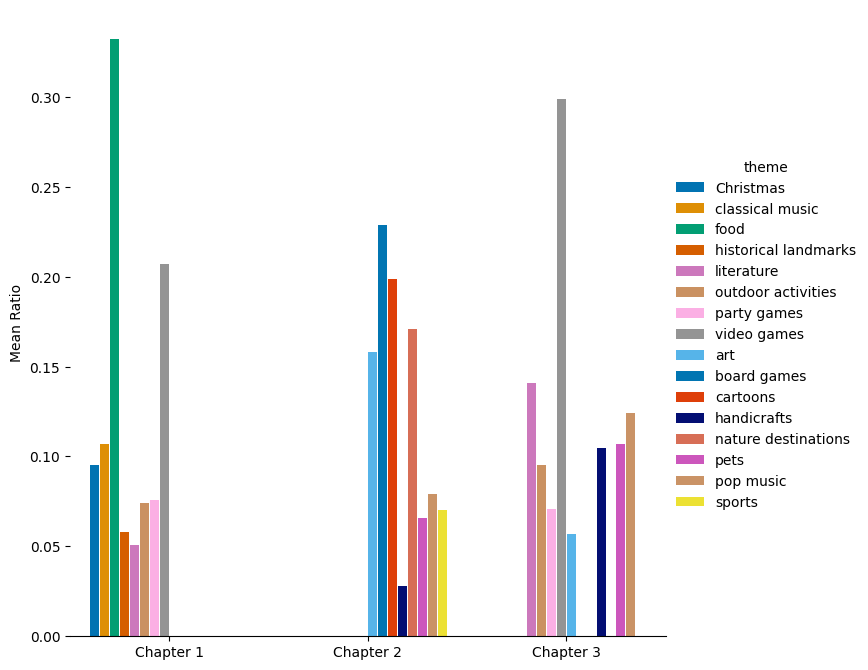}
\caption{Theme popularity per chapter.}
\label{fig:popular_theme}
\end{figure*}

Table \ref{tab:summary_tool} shows that the majority of the recorded interactions with the tool belonged to the first chapter. Since the course was self-paced and with a high attrition rate, there were fewer users in the latter chapters of the course. The tool included the following distinct themes in the first chapter menu: 
\textit{Christmas, classical music, food, historical landmarks, literature, party games, video games}  and \textit{outdoor activities}. Since some of the users were more active than the others, the theme popularity shown in Figure \ref{fig:popular_theme} is the mean ratio of ratios per each individual user. According the Figure \ref{fig:popular_theme}, \textit{food} and \textit{video games} were the most engaging to the course participants. The second and third chapters saw a decline in the user interaction. The second chapter offered the following set of themes: \textit{art, board games, cartoons, handicrafts, nature destinations, pets, pop music,}  and \textit{sports}. In the second chapter, the popular choices were \textit{board games, cartoons, nature destinations} and \textit{art}. The third chapter exhibited the same trend as the second one, which is a further decline in the student engagement with the tool. \textit{Video games}, a well-liked theme in the first chapter, was also the most popular among the users in the third chapter. Table~\ref{tab:random_spec_theme_concept} indicates that the users preferred choosing individual themes over random ones across all the chapters, which is also reflected in their feedback responses. While the context personalization of the exercises was well received by the users, Table \ref{tab:random_spec_theme_concept} indicates that this was not the case for the concept personalization. The study participants displayed a preference for particular concepts in the first chapter; however, they mostly picked random concepts in the second and third chapters. During the expert evaluation, we noticed that the choice of concepts did not appear to significantly affect the variability in the generated exercises, which could have reduced the students' interest in selecting specific concepts . For example, most  of the exercises on ``user input'' also involved ``output'' and vice versa, essentially making the choice between them meaningless.


\begin{table}[h]
\centering
\caption{Comparison of random vs. specific context and concept preferences.}
\label{tab:random_spec_theme_concept}
\begin{tabular}{|c|c|c|c|c|c|} \hline  

 \multicolumn{2}{|c|}{}& \multicolumn{2}{|c|}{\textbf{Theme choice}} & \multicolumn{2}{|c|}{\textbf{Concept choice}} \\ \hline  

\textbf{Chapter} & \textbf{Type} & \textbf{Random} & \textbf{Specific} & \textbf{Random} & \textbf{Specific} \\ \hline  

1 & Fetched & 22\% & 78\% & 35\% & 65\% \\ \cline{2-6}  
 & Solved & 22\% & 78\% & 35\% & 65\% \\ \hline  
2 & Fetched & 42\% & 58\% & 67\% & 33\% \\ \cline{2-6}  
 & Solved & 40\% & 60\% & 47\% & 53\% \\ \hline  
3 & Fetched & 35\% & 65\% & 55\% & 45\% \\ \cline{2-6}   
 & Solved & 41\% & 59\% & 56\% & 44\% \\ \hline 

\end{tabular}
\end{table}

\begin{figure*}[ht]
\centering
\includegraphics[width=0.65\textwidth]{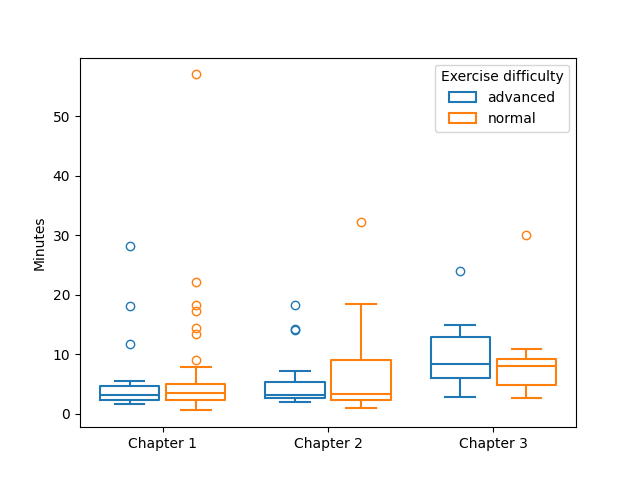}
\caption{Distribution of time between fetching and solving an exercise per each individual user and exercise combination.}
\label{fig:time_per_solve}
\end{figure*}

For the solved exercises, we investigated how much time passed between fetching an exercise and submitting its correct solution. There were 4 cases of the users fetching the same exercise twice but solving it once\footnote{This could happen because skipped exercises were returned to the pool.}. We concluded that the retrieval closest to the submission is the most plausible and discarded the other event. Additionally, there were a few outliers where it took some students over an hour (up to 2 days) to submit their solutions. They were omitted, as it is unlikely those students actually worked on the exercises that long and instead might have taken a break between fetching and solving the exercises. Figure \ref{fig:time_per_solve} shows that the users generally did not require more time for the advanced exercises and on average could solve one problem in less than 10 minutes.

\begin{table*}[h]
\centering
\caption{Summary of exercise results.}
\label{tab:exer_results_combined}
\begin{tabular}{|c|c|c|c|c|c|c|}
\hline
\textbf{Chapter} & \multicolumn{2}{c|}{\textbf{Exercise difficulty}} & \multicolumn{2}{c|}{\textbf{Clarity}} & \multicolumn{2}{c|}{\textbf{Advanced concepts}} \\ \hline
\multicolumn{7}{|c|}{\textbf{Unsolved Exercises}} \\ \hline
 1 & Normal & 66\% & Clear & 96\% & Absent & 83\% \\ \cline{2-7} 
 & Advanced & 34\% & Partially clear & 4\% & Present & 17\% \\ \hline
 2 & Normal & 65\% & Clear & 90\% & Absent & 92\% \\ \cline{2-7} 
 & Advanced & 35\% & Partially clear & 10\% & Present & 8\% \\ \hline
 3 & Normal & 82\% & Clear & 97\% & Absent & 100\% \\ \cline{2-7} 
 & Advanced & 18\% & Partially clear & 3\% & Present & 0\% \\ \hline
\multicolumn{7}{|c|}{\textbf{Solved Exercises}} \\ \hline
 1 & Normal & 66\% & Clear & 98\% & Absent & 100\% \\ \cline{2-7} 
 & Advanced & 34\% & Partially clear & 2\% & Present & 0\% \\ \hline
 2 & Normal & 58\% & Clear & 93\% & Absent & 100\% \\ \cline{2-7} 
 & Advanced & 42\% & Partially clear & 7\% & Present & 0\% \\ \hline
 3 & Normal & 50\% & Clear & 100\% & Absent & 100\% \\ \cline{2-7} 
 & Advanced & 50\% & Partially clear & 0\% & Present & 0\% \\ \hline
\end{tabular}
\end{table*}



\begin{table*}[h]
\centering
\caption{Difficulty match of solved and unsolved exercises.}
\label{tab:difficulty_match}
\begin{tabular}{|c|c|c|c|c|c|c|}
\hline
\textbf{Chapter} & \multicolumn{3}{c|}{\textbf{Solved exercises}} & \multicolumn{3}{c|}{\textbf{Unsolved exercises}} \\ \hline
 & Too easy & Okay & Too difficult & Too easy & Okay & Too difficult \\ \hline
 1 & 22\% & 78\% & 0\% & 24\% & 64\% & 12\% \\ \hline
 2 & 33\% & 65\% & 1\% & 27\% & 63\% & 10\% \\ \hline
 3 & 57\% & 43\% & 0\% & 30\% & 68\% & 2\% \\ \hline
\end{tabular}
\end{table*}


Table~\ref{tab:exer_results_combined} illustrates the differences between the exercises that were solved and unsolved in terms of difficulty, clarity, and presence of advanced concepts. The exercises' difficulty corresponds to the generated difficulty, whereas the other statistics (clarity and presence of advanced concepts) are taken from the expert evaluation. The unsolved exercises contain more occurrences of only partially clear problem descriptions and advanced concepts that were not covered in the course materials. Table \ref{tab:difficulty_match} shows that in the first chapter 12\% of the unsolved exercises were too difficult and 10\% in the second. Additionally, 57\% of the solved exercises in the third chapter consisted of the problems that were too easy for their declared difficulty level.

\section{Discussion}

Both the expert and student evaluations indicate that the quality of the generated exercises was high. In addition, they were well received by the students. This suggests that AI-generated problems could be a valuable addition to introductory programming courses, at least, as additional practice material as was the case in this study. One reason for the positive student feedback could be that the context personalization made the tasks feel more authentic to them, which has been found to positively impact students' perceptions of assessment~\cite{gulikers2006relations,gerritsen2019students,struyven2003students}. These results support earlier findings by Sarsa et al.~\cite{Sarsa_2022}, Jordan et al.~\cite{jordan2024need}, and del Carpio Gutierrez et al.~\cite{DelCarpioGutierrez2024} who also found the quality of generated exercises to be generally high. The quality of the exercises was especially impressive considering that the language of the course was Dart, which is not as popular as many other languages; thus, the model is likely to have seen less Dart code during training compared to more common languages, such as Python, Java, or C.

One potential downside of our approach is that GPT-4 failed to produce content according to the specified difficulty level, e.g., some was too easy or included concepts that were too advanced. Thus, we suggest that this approach is best suited for providing optional practice opportunities that would complement instructor-crafted programming exercises. On the other hand, many shallowly personalized exercises could be solved with practically identical code; consequently this automated approach to exercise generation could be used for circumventing plagiarism. Students could be provided with test problems containing minor variations, as prior work has found little effect of contextualization on programming problem solving performance~\cite{Lovellette2024,Bouvier2016}. Examining the generated exercises from the point of view of the depth and granularity of personalization, our results indicate that the majority (64.0\%) included shallow personalization, especially when the model was severely limited in the number of concepts it could use, as was the case for Chapter 1. Since practically all the exercises were tailored to their narrow topic, we conclude that the granularity of the personalization was fine in the vast majority of the cases. This suggests that our approach is suitable for producing mostly shallow context personalization; however, it is possible to obtain problems of the fine grain size of personalization by providing a model with a specific topic of interest.

The results suggest that using LLMs for contextual personalization of exercises is meaningful for stimulating learners' engagement and potentially preventing boredom, which is often reported by students in computing education~\cite{Coto2021}. In the optional feedback survey (Figure \ref{fig:after_three_exercises}), most of the students agreed that being able to select a theme was enjoyable. What is more, both the expert and student evaluation found that the majority of the exercises matched their declared theme (Figure \ref{fig:after_each_exercise} and Table \ref{tab:expert_eval_results}). The interaction data supported this by showing that the course participants were more likely to choose a specific theme over a random one (Table \ref{tab:random_spec_theme_concept}) and clearly displayed predilections for particular themes (Figure \ref{fig:popular_theme}). Since there is an apparent preference for contextually relevant exercises, providing them could lead to increased engagement and motivation among learners and higher course participant retention~\cite{Guzdial_2010, Guzdial2007, Bouvier2016, Edwards2020}. Improved engagement with exercises, in turn, could lead to higher time-on-task, which has been found to be a good predictor of performance in introductory programming~\cite{leinonen2022time}. Because the exercise tool employed a light gamification element in the form of a trophy given for solving three exercises per chapter, such repetitive engagement could also help students' test scores, as has been demonstrated by Edwards et al.~\cite{Edwards2020}. Moreover, the optional student feedback indicated that the course participants positively evaluated the utility of the exercises for their learning (Figure \ref{fig:after_three_exercises}). This result is in line with earlier research on hands-on approaches to computing education, which have been perceived as beneficial by students~\cite{Ly2021, Sullivan2021}. 

Additionally, contextually personalized exercises may contribute to enhanced situational interest that can potentially lead to the more advanced phases of interest development, i.e., emerging and well-developed individual interest stages~\cite{hidi2006, Michaelis2022, Hoegheim2015, Hoegheim2017, Bernacki2018}. For example, research in mathematics education by Høgheim and Reber~\cite{Hoegheim2017} has indicated that shallow context personalization positively affects students with low individual interest in the subject by supporting their situational interest. Since the majority of the generated problems were shallowly personalized (Table \ref{tab:expert_eval_results}), this effect could potentially translate to computing education students. For those students who have already achieved the initial stage of individual interest development, i.e., emerging individual phase, such exercises can help maintain their interest by serving as an additional source of learning materials. As some of the exercises were personalized to the cultural background of the study participants, they could increase learners' sense of intrinsic and attainment values, as was demonstrated by Schoenherr~\cite{Schoenherr2024}.






\subsection{Limitations}

There are some limitations to this study, which we outline here. First, the course was an elective, self-paced, online course, where the participants were mostly lifelong learning students (i.e., not formally enrolled at a university). Moreover, earlier empirical research on contextually personalized learning materials has been mostly limited to mathematics in secondary education ~\cite{Schoenherr2024, Walkington2017, Walkington2018, Bernacki2018, Hoegheim2015, Hoegheim2017}. Thus, it is uncertain whether the results found here would generalize to more traditional introductory programming courses with deadlines, where the majority of participants are computer science or other STEM majors. Such students might be more motivated to complete such courses in general compared to lifelong learners who constitute the majority of the course population. The experiment was also done at a single institution, which also limits its generalizability.

Another limitation is the lack of background information concerning the preexisting levels of interest and competence in programming among the course participants. Earlier research on context personalization has shown that those students who possess less interest and lower self-perceived competence in a subject, e.g., mathematics, benefit more from contextually personalized problems~\cite{Hoegheim2015, Hoegheim2017}. In their case, context personalization triggers situational interest. What is more, the depth of personalization affects various categories of learners in a different fashion, e.g., those who engage more with mathematics through their interests are more positively influenced by deep personalization in mathematical problems~\cite{Walkington2018}. Since we did not obtain any detailed data on the course participants' prior engagement and attitudes towards computer programming, we could not assess how different levels of personalization affected their study progress or feedback on the exercises. Additionally, we are uncertain to which extent contextualization might have influenced their situational or individual interest in the subject area.

As the course was offered entirely online and there was no contact between the study authors and the study participants, the risk of subject and experimenter expectancy biases~\cite{goebel1971effects,zoble1969interaction} could be considered small due to these effects being primarily evidenced in interpersonal settings ~\cite{harris1985mediation,klein2012low,rosenthal1978interpersonal}. Nonetheless, we acknowledge that, apart from the open feedback, all of our survey questions were Likert-scaled and positively inclined towards effects desired by the authors, e.g., ``The exercises description matched the selected theme'' and ``I enjoyed being able to select themes that match my interests''. This makes our surveys susceptible to an acquiescence bias, where responders tend to passively agree (or disagree) to asked questions irrespective of content~\cite{podsakoff2003common}. However, both age and education have been observed to mitigate the effect of acquiescence on questionnaires~\cite{costello2015acquiescence}, and as the course participants in general are lifelong learners and university students, both the average age and education among the participants was likely to be reasonably high. Besides the phrasing of the questions themselves, the questionnaires or the shown exercises did not contain anything that could be interpreted as suggestive.

Related to the generated exercises, we opted to pregenerate the exercises to avoid running into any problems with an on-demand system, such as the LLM being unavailable for some reason or students trying to do prompt injection attacks to break the tool~\cite{perez2022ignore}. The downside of this approach was that the theme selection was also predefined and thus might not have matched the students' top interests. However, multiple varying themes were available, and we were mostly interested in seeing if students were keen to select any particular theme over a random one. Another limitation of having the exercises pregenerated is that the exercise pool was limited, and consequently the situation did not exactly mimic one where students would have truly unlimited practice opportunities.

Completing the exercises was optional, and no course credit was provided for completing them. This may have caused a selection bias, as active students might have used the tool more frequently. Since we openly told the course participants that the exercises were AI-generated, it is also possible that those students who are interested in AI were more likely to complete them, potentially skewing the results of the surveys. Such students might have also rated the AI-generated content more favorably compared to those learners who have more neutral or negative views on AI.

Related to the expert evaluation, two questions had relatively low agreement scores: determining if the exercise difficulty matched the selected difficulty and whether the personalization was deep or shallow. Thus, the results utilizing the expert evaluation data for these two questions should be taken with a grain of salt -- the other evaluators could have made different decisions on these questions.

One limitation of the experimental design was that the course was organized in Finnish, but the tool, surveys, and generated exercises were in English. We opted to generate exercises in English as prior work has found that LLMs perform the best in English~\cite{zhao2024llama}. Recent work exploring the generation of exercises with LLMs in other languages besides English has demonstrated promising results~\cite{jordan2024need}, which is why we are interested in implementing the tool in Finnish in the future. The language of the exercise descriptions could have affected students' ratings as they are not native English speakers; however, Finns generally possesses good English skills.








\section{Conclusions}

In this work, we explored how successfully LLMs, namely GPT-4, can generate programming exercises. The generated content was offered for additional student practice in an online programming course. We evaluated the quality of the generated exercises by having both the authors and the students in the course assess them. In addition, we examined how the course participants interacted with the exercises in the course e-book.

Our findings indicate that the vast majority of the exercises generated by the LLM were clear and matched various themes, topics, and concepts. What is more, they rarely included concepts that were too advanced for the students in the course. However, the exercises were often easier than requested in the prompt given to the model, and the thematic personalization was often shallow.

In our future work, we are interested in studying how adding gamification features to the tool affects student engagement. Additionally, we are interested in examining how students' programming experience correlates with their interaction with LLM-generated exercises. We are working on a version of the tool that would allow students to generate exercises related to any contextual theme on demand, instead of having a list of predefined themes available in the tool.

\begin{acks}
This research was supported by the Research Council of Finland (Academy Research Fellow grant number 356114).
\end{acks}


\balance
\bibliographystyle{ACM-Reference-Format}
\bibliography{references}


\end{document}